\documentclass[aip,jcp,reprint]{revtex4-1}
   \usepackage{amsmath}   %
   \usepackage{float}   %
   \usepackage{bm}        % bold math symbols
   \usepackage{amsfonts}  % revtex parameters for fonts, symbols etc.
   \usepackage{amssymb}   %
   \usepackage{graphicx}  % Include figure files
   \usepackage{todonotes}
   \usepackage{array}
   \usepackage{braket}
   \usepackage{color}
\usepackage[colorlinks=true,linkcolor=blue, citecolor=blue, urlcolor=blue]{hyperref}
\usepackage{url} %package for including web-links
\usepackage[all]{hypcap} % allows references to jump to the beginning of the figure, not to the caption
\usepackage{multirow} % multi-row entries in tables

\newcolumntype{L}[1]{>{\raggedright\let\newline\\\arraybackslash\hspace{0pt}}m{#1}}
\newcolumntype{C}[1]{>{\centering\let\newline\\\arraybackslash\hspace{0pt}}m{#1}}
\newcolumntype{R}[1]{>{\raggedleft\let\newline\\\arraybackslash\hspace{0pt}}m{#1}}

\renewcommand{\thispagestyle}[1]{}

\begin{document}
\title{Global Structure Search for Molecules on Surfaces: Efficient Sampling with Curvilinear Coordinates}
% \begin{flushright}submitted to J. Chem. Phys. \end{flushright}
\author{Konstantin Krautgasser}
\affiliation{Department Chemie, Technische Universit{\"a}t M{\"u}nchen, Lichtenbergstr. 4, D-85748 Garching, Germany}
\author{Chiara Panosetti}
\affiliation{Department Chemie, Technische Universit{\"a}t M{\"u}nchen, Lichtenbergstr. 4, D-85748 Garching, Germany}
\author{Dennis Palagin}
\affiliation{Physical and Theoretical Chemistry Laboratory, Department of Chemistry, University of Oxford, South Parks Road, Oxford, OX1 3QZ, United Kingdom}
\author{Karsten Reuter}
\affiliation{Department Chemie, Technische Universit{\"a}t M{\"u}nchen, Lichtenbergstr. 4, D-85748 Garching, Germany}
\author{Reinhard J. Maurer}
\email[]{reinhard.maurer@yale.edu}
\affiliation{Department of Chemistry, Yale University, New Haven, CT 06520, USA}

\date{\today}

\begin{abstract}
Efficient structure search is a major challenge in computational materials science. We present a modification of the basin hopping global geometry optimization approach that uses a curvilinear coordinate system to describe global trial moves. 
This approach has recently been shown to be efficient in structure determination of clusters [Nano Letters 15, 8044--8048 (2015)] and is here extended for its application to covalent, complex molecules and large adsorbates on surfaces. The employed automatically constructed delocalized internal coordinates are similar to molecular vibrations, which enhances the generation of chemically meaningful trial structures. By introducing flexible constraints and local translation and rotation of independent geometrical subunits we enable the use of this method for molecules adsorbed on surfaces and interfaces. For two test systems, \emph{trans}-$\beta$-ionylideneacetic acid adsorbed on a Au(111) surface and methane adsorbed on a Ag(111) surface, we obtain superior performance of the method compared to standard optimization moves based on Cartesian coordinates.
\end{abstract}

% \keywords{Basin Hopping, Global Optimization, organic adsorbates, delocalized internals, Density Functional Tight-Binding}

\maketitle 

\section{Introduction}
\label{intro}

In just 20 years the complexity of systems studied in surface science has increased by orders of magnitude.~\cite{Dabrowski_review_2001} Whereas previous major problems where associated with the adsorption behavior of small adsorbates such as CO on Pt(111),~\cite{Feibelman2001}
typical challenges that dominate surface science nowadays are associated with the structural, electronic, and chemical properties of large organic adsorbates and molecular networks on metal and semiconductor surfaces.~\cite{Barlow_Raval_2003, Barth_review_2007, Wolkow1999} Here the interaction strength and geometry of systems and the implications for nanoarchitectures, self-assembly, or atomic-scale manipulation is studied by two complementary sets of techniques, namely by \emph{in vacuo} surface science experiments~\cite{surface_science_exp_book} and \emph{ab initio} electronic structure simulations.~\cite{Hammer_Norskov_2000} The capabilities of both experiment and simulation have increased tremendously. On the computational side,  previously unthinkable system sizes can be dealt with thanks to the immense increase in computational power and the numerical, as well as algorithmic efficiency of Density-Functional Theory (DFT).~\cite{Becke_DFT_review_2014} Owing to the latest advances in particular in the treatment of dispersive interactions, such calculations are nowadays able to provide a reliable description of the adsorption structure and energetics of highly complex organic adsorbates on metal surfaces.~\cite{Maurer2016, Mercurio2013, Maurer2016Review}

An especially important aspect of theoretical surface science that should benefit from this progress is the ability to conduct computational screenings of possible geometries of complex interfaces, be it for determining the atomic configuration of surface reconstructions,~\cite{surface_reconstruction_screening} topologies of organic crystals,~\cite{crystal_structure_prediction} or identifying the manifold of all possible compositions and adsorption sites in processes of heterogeneous catalysis and in compound   materials.~\cite{Norskov_screening, Hutchison_screening, Henning_screening} However, one thing that has not changed much in recent years is how, in simulations, optimal adsorption structures are identified. The large number of degrees of freedom in a complex adsorbate, as well as the structural complexity of large surface nanostructures, necessitates a full global search for optimal structures and overlayer phases that includes all possible binding modes and chemical changes that can occur upon adsorption. Knowing the geometric structure of an interface is an important prerequisite to investigating its electronic properties and level alignment, which in turn determines the performance in potential electronic or catalytic applications. An efficient potential energy surface (PES) sampling tool is necessary to enable this, as opposed to just conventional local optimization of several chemically sensible initial guesses. In general, this problem can be solved by applying global geometry optimization methods.~\cite{Wales_Doye_2000_review, global_opt_book_2006}

The goal of a majority of such global optimization approaches is to efficiently traverse the PES by generating new trial configurations that are subsequently accepted or rejected based on certain criteria. Multiple reviews on global optimization methods specific to certain types of systems, ranging from small clusters to large biomolecules, have been published recently.~\cite{Wales_1998_Nature, Wales_1999_Science} In particular, much attention has been attracted by the application of global optimization methods to the study of metal clusters,~\cite{Johnston_2003_review} binary nanoalloys,~\cite{Ferrando_Johnston_2008_review} proteins,~\cite{Scheraga_proteins_review_1999} water clusters,~\cite{Hartke_water_global} and molecular switches.~\cite{Carstensen_Hartke_2011} Recently, global geometry optimization schemes have been specifically developed and applied in the growing field of heterogeneous catalysis,~\cite{Peterson2014} including application in reaction coordinate prediction,~\cite{Wales_2015_perspective} and biomolecular simulations.~\cite{Blum2015} A wide range of different global optimization algorithms has thereby been suggested over the last two decades, ranging from simple classical statistical mechanics simulated annealing schemes~\cite{Kirkpatrick1983} to sophisticated landscape paving,~\cite{landscape_paving} puddle-jumping,~\cite{puddle_skimming, puddle_jumping} or neural-network controlled dynamic evolutionary approaches.~\cite{Styrcz_2011, Hase2016} Two prominent and most popular families of global geometry optimization techniques include Monte-Carlo based methods, such as basin hopping (BH),~\cite{Wales1997} and evolutionary principles based genetic algorithms (GA).~\cite{Deaven_1995_GA} However, no general rule of preference to a specific algorithm exists, as the efficiency of classical global optimization methods is highly system dependent.~\cite{Ferrando_Johnston_2008_review}

There are many technical aspects that influence the performance of any global sampling technique, such as the choice of initial geometry, the ways of disturbing the configuration during the trial move, the definition of acceptance criteria, the methods employed to calculate potential energies and forces \emph{etc}. Many authors were concerned with the efficiency of applied sampling methods and suggested various improvements.~\cite{Gehrke_Reuter_2009, Goedecker_JCP_2009, Goedecker_JCP_2011, Lai_Xu_Huang_JCP_2011, Lyakhov_2013, Rondina_2013} Moreover, the possibility of moving from the total energy as the main sampling criterion towards observable-driven~\cite{Hartke_observable_targeting} and grand-canonical~\cite{Wales_grand_canonical} global sampling schemes has been suggested. The parameter that plays the most crucial role in method performance, however, is the choice of coordinates suitable for representing the geometries and, most importantly, changes in geometries during the sampling, \emph{i.e.}\;the so-called trial moves.~\cite{Niesse_Mayne_JCP_1996} The essential importance of the trial geometry generation step has already been noticed in early system-specific publications.~\cite{Niesse_Mayne_JCP_1996, Iwamatsu_2000} For instance, simple group rotations within proteins\cite{Wales_2014} or clusters\cite{Wales1997} already led to significant gains in sampling efficiency. Furthermore, several computationally more expensive ways to produce elementary trial moves were suggested based on short high-temperature molecular dynamics (MD) runs.\cite{Ferrando_2009, Goedecker_JCP_2009} For certain systems they have indeed proven to be much more effective, allowing to \emph{e.g.} optimize large metallic clusters within atomistic models.~\cite{Ferrando_ACSNano_2008, Ferrando_NanoLett_2010} 

From a programmer's point of view, the perhaps most intuitive representation of the atomic coordinates of trial moves are Cartesian coordinate (CC) displacements. It has been noted though that this most popular representation is often inefficient,~\cite{Niesse_Mayne_JCP_1996, Hartke_JPC_1993} due to its chemical blindness, which may easily lead to unphysical configurations (\emph{e.g.}\;dissociated structures). CC displacements do not account for the overall geometry and the coupling between coordinates. Different approaches have been proposed to overcome this difficulty, such as virtual-alphabet genetic algorithms,~\cite{Goldberg_1991} employing the idea that coordinates should stand as meaningful building blocks,~\cite{Goldberg_1989} or genetic algorithms with space-fixed CCs,~\cite{Niesse_Mayne_JCP_1996, Zeiri_1995, Iwamatsu_2000} introducing non-traditional genetic operators. The main limitation of these methods, however, is that they remain rather system specific. The choice of suitable coordinates for global optimization should rather be system independent, while at the same time adapted to the chemical structure. Especially in the case of large adsorbates the completely unbiased way in which the structural phase space is sampled in CCs does not allow for an efficient structural search for conformational changes upon adsorption that leave the molecular connectivity intact. The chemically most intuitive set of coordinates for such a situation would instead be internal coordinates (ICs), namely bond stretches, angle bends, or torsions. Already in the seminal work on BH of Wales and Doye, the potential usefulness of such coordinates has been noted,~\cite{Wales1997} and later shown to be beneficial for structures connected by double-ended pathways.\cite{Wales2012}

One of the first applications of the idea of using ICs for global geometry optimization was reported in the context of protein-ligand docking, which developed into the so-called Internal Coordinate Mechanics (ICM) model,~\cite{Abagyan_docking_2002} further extending the above mentioned concept of meaningful building blocks.~\cite{Goldberg_1989} Another example of global optimization in ICs was the attempt of introducing dihedral angles into the framework of the so-called deterministic global optimization.~\cite{Floudas_JCP_1994} Global optimization of clusters and molecules in ICs using the Z-matrix representation was suggested by Dieterich and Hartke.~\cite{OGOLEM_code_reference} However, automatic construction of a Z-matrix is close to impossible for larger, more complicated systems, especially organic molecules containing rings. Besides, it also does not eliminate the problem of coordinate redundancies, which generally limits the applicability of this approach.

Especially suitable for a system independent description of complex molecular structures are instead so-called delocalized internal coordinates (DICs), \emph{i.e.} non-redundant linear combinations of ICs, that have been extensively used for efficient local structure optimization of covalent molecules,~\cite{VonArnim1999, Pulay1992, Baker1996, Baker1999} crystalline structures,~\cite{Bucko2005,Jahnatek2009,Egger2014} and for vibrational calculations.~\cite{Piccini2014, Strobusch2013} In our recent work~\cite{Panosetti2015}, we implemented such automatically generated collective curvilinear coordinates in a BH global sampling procedure. The similarity of these coordinates to molecular vibrations does yield an enhanced generation of chemically meaningful trial structures, especially for covalently bound systems. In the application to hydrogenated Si clusters, we correspondingly observed a significantly increased efficiency in identifying low-energy structures when compared to CC trial moves and exploited this enhancement for an extensive sampling of potential products of silicon cluster soft landing on the Si(001) surface.

In the present work, we provide a detailed methodological account of this curvilinear coordinate global optimization approach,~\cite{Panosetti2015} and extend it to a conformational screening of adsorbates on surfaces. We do this by introducing constraints and extending the coordinate system to include lateral surface translations and rigid adsorbate rotations. 
These curvilinear coordinates are constructed automatically at every global optimization step and, similarly to the original BH scheme, we pick a random set of coordinate displacements with which a trial move is attempted. Testing this for the \emph{trans}-$\beta$-ionylideneacetic acid molecule adsorbed on a Au(111) surface (as an example of a complex functionalized organic adsorbate), and for methane adsorbed on a Ag(111) surface (as an example of a small adsorbate with a shallow PES and many minima), we find that DIC trial moves increase the efficiency of BH by both a more complete sampling of the possible surface adsorption sites and by reducing the number of molecular dissociations.

The paper is organized as follows: In Chapter~\ref{methods} we outline traditional approaches to global structure optimization, explain the construction of DICs and our definition of a complete coordinate set for adsorbates on surfaces, and how these coordinates are applied for conformational structure searches in a global optimization framework such as basin hopping. In Chapter~\ref{computational} we summarize the computational details of our benchmark studies on the surface-adsorbed aggregates presented in Chapter~\ref{results}. We conclude our work in Chapter~\ref{conclusions}. Additional remarks on software and algorithms can be found in Appendix~\ref{appendix-winak}, details on DFT calculated diol stability can be found in Appendix~\ref{appendix-dft}.

\section{Methods}
\label{methods}

\subsection{Global Optimization - Basin Hopping}

A variety of global optimization procedures exists to date, including simulated annealing,~\cite{Kirkpatrick1983} genetic algorithms,~\cite{Deaven_1995_GA} BH,~\cite{Wales1997, Wales_Doye_2000_review} and many variants and combinations thereof.~\cite{Gehrke_Reuter_2009, Goedecker_JCP_2009, Goedecker_JCP_2011} One reason for the success of basin hopping might be the intriguing simplicity of the approach, summarized as follows:
\begin{enumerate}
 \item Displace a starting geometry with a random \emph{global} Cartesian coordinate (CC) trial move $\Delta \mathbf{x}$ normalized to a step width $dr$;
 \item Perform a \emph{local} structure optimization to a minimum energy structure with energy $E_i$;
 \item Pick a random number and accept the new minimum energy structure with the probability
 \begin{equation}
  P(\Delta E_i)=\exp\left(-\frac{E_i-E_{\mathrm{min}}}{k_B T_{\mathrm{eff}}}\right) \quad ,
 \end{equation}
 where $T_{\mathrm{eff}}$ corresponds to an effective temperature and $E_ {\mathrm{min}}$ refers to the energy of the current lowest energy structure;
 \item If the structure is accepted, place it as a new starting point and proceed with point 1.
\end{enumerate}
Following this procedure, as illustrated in Fig. 1, the PES is sampled until all chemically relevant minima are found. A sufficiently high $T_{\mathrm{eff}}$ ensures that also higher-lying minima are accepted with adequate probability and a large area of the PES can be sampled. The size of trial moves $\Delta\mathbf{x}$ must be large enough to escape the basin of attraction of the current local minimum, but small enough not to lead to regions of the PES that are of little interest, \emph{e.g.} dissociated structures or structures with colliding atoms. The corresponding displacement $\Delta\mathbf{x}$ is constructed from random Cartesian vectors normalized to the chosen step width. This approach has proven highly efficient for optimization of clusters and biomolecules.~\cite{Wales_1999_Science, Palagin2011} In the following we will discuss a  modification of this general algorithm only with respect to point 1 by changing the construction of the displacement vector $\Delta\mathbf{x}$.~\cite{Panosetti2015}

\begin{figure}[h]
\centering
 \includegraphics[width=\columnwidth]{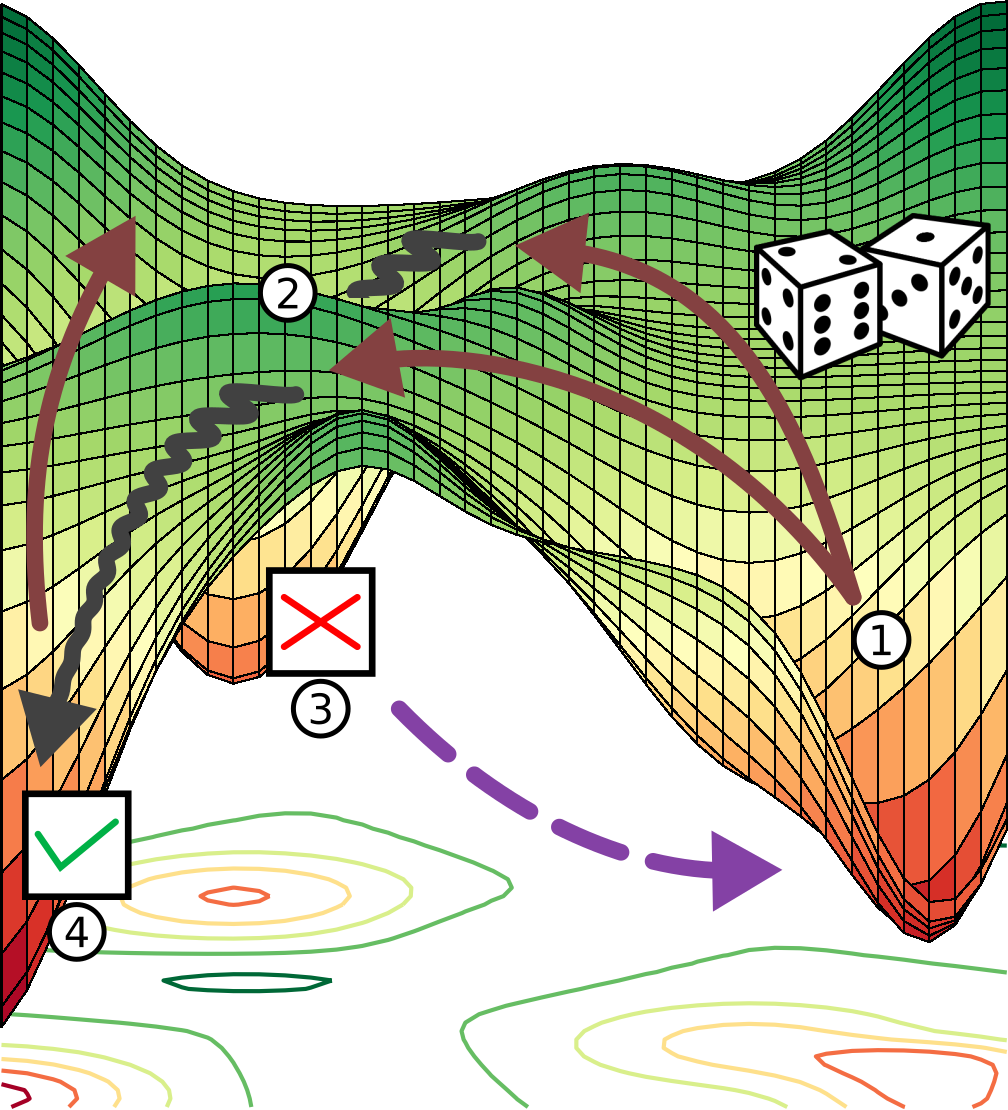}
\caption{\label{fig-bh}Illustration of the basin hopping global optimization approach. Random trial moves $\Delta\mathbf{x}$ (straight arrows, 1) displace the geometry towards a new basin of attraction and local optimization (zigzag arrows, 2) identifies the corresponding minimum energy structure. The identified minima will be rejected (3) or accepted (4) with a certain probability given by eq. 1.}
\end{figure}

\subsection{Construction of Delocalized Internal Coordinates}

ICs can be defined in many different ways. A typically employed approach is to define a set of coordinates in a body-fixed frame as a set of bond stretches, angle bends, torsional modes, and out-of-plane angles.~\cite{Wilson1980} Any displacement $\Delta\mathbf{x}$ from a given point $\mathbf{R}_{\mathbf{x}_0}$ in Cartesian coordinate space can then be transformed to a set of internal coordinate displacements $\Delta\mathbf{q}$ by
\begin{equation}
 \Delta\mathbf{q} = \mathbf{B} \Delta\mathbf{x} \quad ,
\end{equation}
where 
\begin{equation}
  \mathbf{B} = \frac{d\mathbf{q}}{d\mathbf{x}} \quad .
\end{equation}
The set of coordinates in $\mathbf{q}$-space can be chosen to be non-orthogonal and complete ({\em e.g.} via a Z-matrix), however this is often difficult and highly system specific. 

An alternative approach is the construction of a highly redundant set of $M>(3N-6)$ primitive ICs to represent the independent degrees of freedom of an $N$ atom  system.~\cite{Peng1996} Pulay, Fogarasi, Baker, and others~\cite{Pulay1979, Pulay1992, Baker1996, Baker1999, Peng1996,Piccini2014} have shown that an orthogonal, complete set of coordinates can be constructed from such a redundant coordinate set simply by singular value decomposition (SVD) of the matrix $\mathbf{G}=\mathbf{B}^\dagger \mathbf{B}$. By construction, matrix $\mathbf{G}$ is positive semi-definite and Hermitian, and satisfies
\begin{equation} \label{eq-svd}
 \mathbf{G} = \mathbf{B}^\dagger \mathbf{B} = \mathbf{U} \left[ \begin{matrix}
  \mathbf{\Lambda} & 0 \\
  0 & 0
 \end{matrix} \right] \mathbf{U}^\dagger \quad .
\end{equation}
In Eq.~\ref{eq-svd}, $\mathbf{U}$ is a set of vectors in the space of primitive ICs $\mathbf{q}$ that defines linear combinations thereof. SVD yields a subset of $(3N-6)$ such independent vectors with non-zero eigenvalues $\mathbf{\Lambda}$ that constitute a fully orthogonal system of DICs.~\cite{Baker1996} These coordinates span all primitive ICs that were previously defined in $\mathbf{B}$.
$\mathbf{U}$ is therefore nothing else than a Jacobian that transforms displacements in the space of primitive internals $\mathbf{q}$ into displacements in the space of DICs $\mathbf{d}$
\begin{equation} \label{eq-transformation}
 \Delta\mathbf{d} = \mathbf{U}\Delta\mathbf{q} = \underbrace{\mathbf{UB}}_{\tilde{\mathbf{B}}} \Delta\mathbf{x} = \tilde{\mathbf{B}}\Delta\mathbf{x} .
\end{equation}

In the last equation we have defined the transformation matrix $\tilde{\mathbf{B}}$ of dimension $(3N-6) \times 3N$ that transforms CC into DIC displacements. Unfortunately the back transformation cannot simply be done by inversion as the matrix $\tilde{\mathbf{B}}$ is non-quadratic and therefore singular. However, we can define a generalized inverse~\cite{Baker1999} that connects displacements in CC and DIC space
\begin{equation}\label{eq-backtransformation}
 \Delta\mathbf{x} = \tilde{\mathbf{B}}^{-1}\Delta\mathbf{d} ,
\end{equation}
where 
\begin{equation}
 \tilde{\mathbf{B}}^{-1} = \tilde{\mathbf{B}}^T\underbrace{(\tilde{\mathbf{B}}\tilde{\mathbf{B}}^T)^{-1}}_{\tilde{\mathbf{G}}^{-1}} .
\end{equation}
Equation~\ref{eq-backtransformation} can easily be verified by multiplication from the left with $\tilde{\mathbf{B}}$. Utilizing the above expressions we have thus established a bijective transformation between the displacements $\Delta\mathbf{x}$ and $\Delta\mathbf{d}$.
However, transformation of the absolute positions of atoms in a molecule, their nuclear gradients and second derivatives requires iterative backtransformation from DIC to CC space. This is due to the fact that $\tilde{\mathbf{B}}$ is simply a linear tangential approximation to the curvilinear hyperplane spanned by coordinates $\mathbf{d}$. Further details on DIC transformation properties can be found in Ref.~\onlinecite{Baker1999}.

In summary, the here presented construction allows to define displacements in terms of a random set of DICs $\Delta\mathbf{d}$ and expression thereof as CC displacements $\Delta\mathbf{x}$, which can then be used to generate a new trial structure from a starting geometry (\emph{cf.} basin hopping procedure, point~1).

\subsection{Curvilinear Coordinates for Adsorbates and Aggregates}
\label{methods-cdics}

\begin{figure}
\includegraphics[width=\columnwidth]{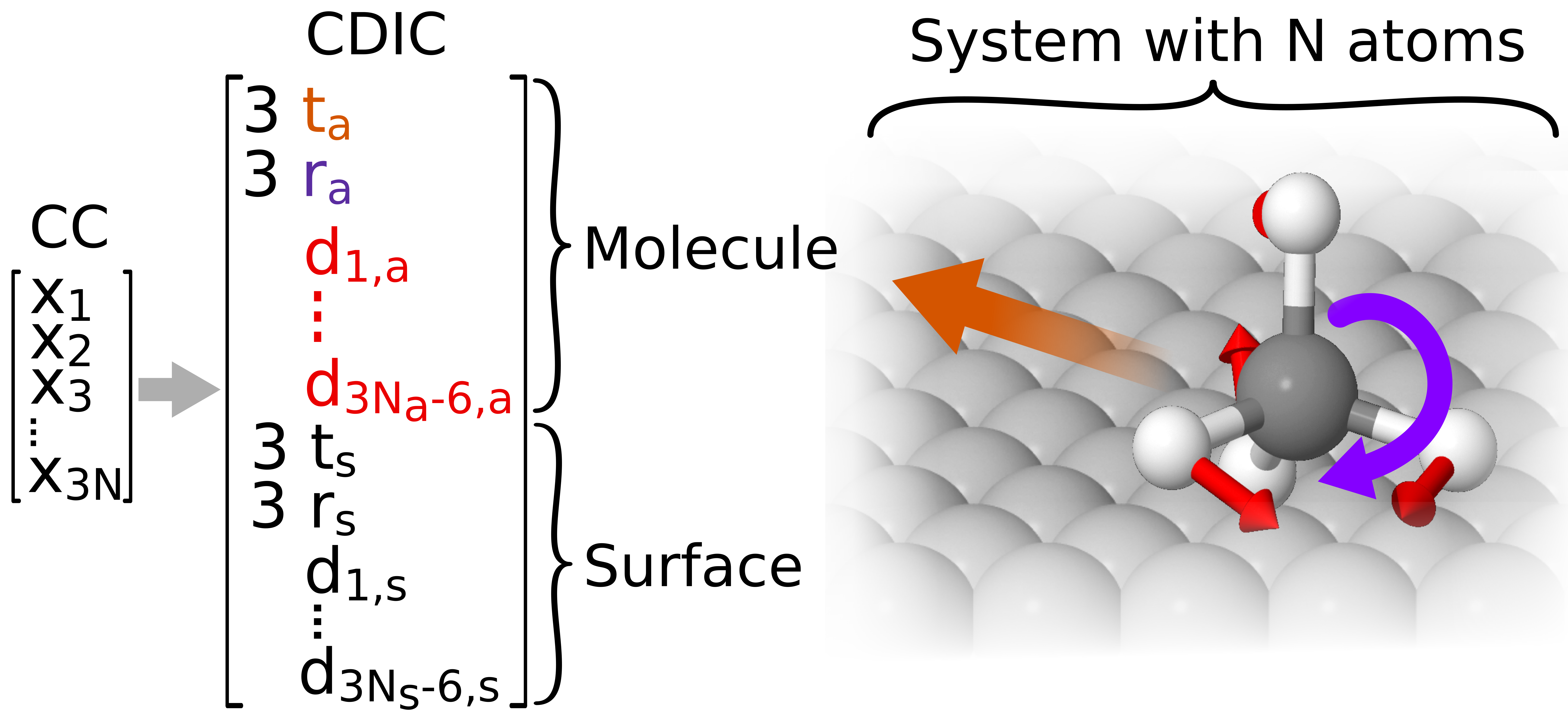}
\caption{\label{fig-CDC-scheme} Illustration of the coordinate definition for a set of complete delocalized internal coordinates (CDIC). Curvilinear coordinates are constructed by partitioning the system into subsystems such as molecule and surface (here shown for methane adsorbed on Ag(111)) and associating each of those with a set of translations (depicted in orange), rotations (purple), and DICs (red). $\mathbf{t}$, $\mathbf{r}$, $\mathbf{d}$ refer to translation, rotation, and DIC vectors, while indices a and s refer to the adsorbate and the surface subsystem, respectively. $N$, $N_{\mathrm{a}}$, and $N_{\mathrm{s}}$ refer to the number of atoms in the complete system or respective subsystem.}
\end{figure}

In the above presented formalism we have constructed $(3N-6)$ DICs, which fully describe all (body-fixed) deformations of a molecule. By construction, overall translations and rigid rotations of isolated molecules and clusters are removed due to their invariance with respect to these coordinates. Re-introducing these degrees of freedom can in some cases be useful, for example in the description of molecules or clusters adsorbed on surfaces, in dense molecular arrangements, or organic crystals. In such systems the position and orientation of a tightly-bound subsystem, such as a molecule, with respect to the rest of the system, such as a surface or the molecular neighbors, is determined by overall weaker interactions and forces.

One can account for this specific chemistry of the system by partitioning it into two (or more) subsystems (\emph{e.g.} adsorbate and surface), for each of which one constructs a set of 3 rigid translations, 3 rigid rotations, and $(3N_{\mathrm{a}}-6)$ DICs, where $N_{\mathrm{a}}$ is the number of atoms in subsystem a. In the following we will refer to such a set of coordinates as \emph{complete} delocalized internal coordinates (CDICs). This idea is illustrated in Fig.~\ref{fig-CDC-scheme}. The pictured system, namely methane on a Ag(111) surface, is partitioned into molecule and surface, and each of these two subsystems is assigned its own set of coordinates, the concatenation of which comprises a full-dimensional coordinate vector that is equivalent to CCs. The translations can be added as simple Cartesian unit vectors of the center of mass (or any set or subset of atoms), as symmetry-adapted linear combinations thereof, or even as fractional CCs in a unit cell, for example to adapt to the symmetry of a given surface unit cell. For the definition of sub-system rotations we have chosen to employ nautical or Tait-Bryan angles. Hereby the three rotation angles $\Phi$, $\Theta$, $\Psi$ (also called yaw, pitch, and roll) are defined around three distinct rotation axes with respect to a reference orientation. This reference orientation is given by the molecular geometry from which the DICs are constructed and establishes an Eckart frame. Changes to orientation are identified and applied by standard rigid-body superposition using Quaternions.~\cite{Kneller1991}

In the systems where the above partitioning makes sense chemically, the concomitant decoupling of coordinates may be beneficial for the description of dynamics and global optimization. In cases where, for example, no significant surface reconstruction is expected upon adsorption, it may not be useful to construct DICs for the substrate, where chemical bonding is more isotropic. Instead, CDIC partitions can be constructed for arbitrary subsets of coordinates (\emph{e.g.} the adsorbate) in order to then be combined with the remaining CCs (\emph{e.g.} the metal substrate). Finally, CDICs (as well as DICs) make the definition of arbitrary coordinate constraints fairly straightforward as is shown in subsection~\ref{section-constraints}.

\subsection{Constraints on Delocalized Internal Coordinates}
\label{section-constraints}
Starting from a set of DICs, constraints on primitive internals such as bond stretches, angle bends, or torsions can be applied by projection from the initial set of DIC vectors $\mathbf{U}$.~\cite{Baker1996} The constraint vector $\mathbf{C}_i$ is taken as a simple unit vector in the space of primitive ICs $\mathbf{q}$
\begin{equation}
 \mathbf{C}_i = \left(
 \begin{smallmatrix}
  0 \\ 
  0 \\ 
  \vdots \\[1ex]
  1 \\ 
  \vdots \\[1ex]
  0 \\
  0 \\
 \end{smallmatrix}
 \right) \quad .
\end{equation}
Correspondingly, the projection of this constraint vector onto the eigenvector matrix $\mathbf{U}$
\begin{equation}
 \tilde{\mathbf{C}}_i=\sum_j\braket{\mathbf{C}_i|\mathbf{U}_{j}}\mathbf{U}_{j}
\end{equation}
yields a vector $\tilde{\mathbf{C}}_i$ in the space of DICs $\mathbf{d}$ (denoted by the tilde sign). All vectors $\mathbf{U}_j$ need to be orthogonalized with respect to $\mathbf{C}_i$ ({\em e.g.} using Gram-Schmidt orthogonalization) to ensure that the primitive coordinate $i$ is removed from the DICs. The resulting matrix $\mathbf{U}'$ is then used to transform back to CCs using the inverse of  $\tilde{\mathbf{B}}'=\mathbf{B}\mathbf{U}'$ in the same way as in Eq.~\ref{eq-backtransformation}.

Constraints on sub-system translations or rotations in CDICs can be applied by simply removing them from the available set of coordinates in $\mathbf{d}$-space. The iterative nature of the backtransformation to absolute CCs may require an iterative coordinate constraint algorithm such as SHAKE.~\cite{Ryckaert1977} Cartesian constraints can be applied by simply excluding atoms or CCs from the reference geometry from which the DICs are constructed, or by projections in Cartesian space $\mathbf{x}$ equivalent to the above projections in $\mathbf{q}$-space.~\cite{Andzelm2001}

\subsection{Global Optimization With Curvilinear Trial Moves}

Summarizing the above, we have defined a set of complete, orthogonal curvilinear coordinates describing all internal motions of molecules, which can be constructed in a fully automatic manner from a set of CCs. For systems with multiple molecules, aggregates, or for surface-adsorbed molecules we can supplement these with translational and rotational degrees of freedom for the individual subsystems. Any changes or displacements applied in these coordinates can be transformed back to CCs and be readily applied in this space.

Displacements in form of DICs or CDICs yield molecular deformations that are more natural to the chemistry of the system, simply due to their construction based on the connectivity of atoms. We have recently shown that application of such coordinates in the framework of global optimization methods can largely facilitate a structure search by preferential bias towards energetically more favorable geometries.~\cite{Panosetti2015} Modification of a global optimization algorithm in terms of DIC trial moves on the example of the BH procedure sketched above can be summarized as follows:
\begin{enumerate}
 \item Displace a starting geometry with a random DIC (CDIC) trial move $\Delta\mathbf{d}$. Therefore:
 \begin{itemize}
  \item Construct a set of DICs (CDICs);
  \item Apply constraints, if necessary;
  \item Pick a random (sub-)set of DICs (CDICs) and normalize them to a given step width $dr$;
  \item Transform coordinates back to Cartesian space using Eq.~\ref{eq-backtransformation} and apply them to the structure;
 \end{itemize}
 \item Perform a \emph{local} structure optimization to a minimum energy structure with energy $E_i$;
 \item Pick a random number and accept the new minimum energy structure with the probability
 \begin{equation}
  P(\Delta E_i)=\exp(-\frac{E_i-E_{\mathrm{min}}}{k_B T_{\mathrm{eff}}}).
 \end{equation}
 \item If the structure is accepted, place it as a new starting point and proceed with point~1.
\end{enumerate}

\section{Computational Details}
\label{computational}

We consider three test sytems: Retinoic acid (ReA, \emph{cf.} Fig.~\ref{fig-gas-pics}) in gas phase; \emph{trans}-$\beta$-ionylideneacetic acid~\cite{Shirley1985} ($\beta$-acid, \emph{cf.} Fig.~\ref{fig-beta}) adsorbed on a six-layer, $(5 \times 5)$ surface unit cell Au(111) slab; and methane (CH$_4$) on a four-layer, $(2 \times 2)$ surface unit cell Ag(111) slab. All test systems were built using the ASE (Atomic Simulation Environment) package.~\cite{Bahn2002} Energetics for the global optimization runs were calculated with Density Functional-based Tight-Binding (DFTB) performed with the DFTB+ V1.1 code.~\cite{Aradi2007, Elstner1998} To correctly account for surface-molecule interactions in the $\beta$-acid on Au(111) case, the Tkatchenko-Scheffler screened dispersion correction method vdW$^{\mathrm{surf}}$~\cite{Ruiz2012} was applied as an \emph{a posteriori} correction to the DFTB results. Since vdW$^{\mathrm{surf}}$ was not available in DFTB+ at the time, we have neglected the environment-dependent rescaling of atomwise dispersion coefficients. Environment-dependence of dispersion interactions in conjunction with tight-binding methods has been introduced recently.~\cite{Stoehr2016} As convergence criterion for the self-consistent DFTB charges, 10$^{-7}$ electrons was used for all systems. DFTB parameters for interaction between C, H, O, Au, and Ag have been generated as described in Refs.~\onlinecite{Michelitsch2014,Makinen2013} and are available upon request. Si and H parameters are described in Ref.~\onlinecite{si_para}. We have used the refinements as described in Ref.~\onlinecite{si_para_impr}.

We implemented the coordinate generation as well as a BH procedure in a modular python package (\texttt{winak}, see Appendix~\ref{appendix-winak}).~\cite{winak} The set of redundant internal coordinates was constructed by including bond lengths, bending angles, and dihedral angles. Bond lengths were included if they were within the sum of the covalent radii of the atoms plus 0.5~\AA{}. Bending angles were included if the composing atoms were bonded to each other and if the bending angle was below 170$^{\circ}$. Dihedral angles were included if the composing atoms were bonded and if the dihedral angle was smaller than   160$^{\circ}$. The effective temperature in the BH procedure was set to 300~K. Trial moves in unconstrained DICs/CDICs and with all stretches constrained (constr.\,DICs/CDICs) were always constructed using linear combinations of a varying percentage of all modes, \emph{e.g.} 10\%, 25\%, 75\% or 100\% of all available coordinates. These subsets were randomly chosen and in the construction of the displacement vector each DIC mode was scaled with a random factor in the range $[-1;1]$. CC displacements were created by assigning a random value in the range $[-1;1]$ to the $x$, $y$ and $z$ coordinates of every atom. As a normalization, all displacements were divided by the single largest component in CCs, such that the maximum absolute displacement of an atom in $x$, $y$, or $z$ direction was 1~\AA{}. Subsequently the whole displacement vector was multiplied by a tunable dimensionless factor \emph{dr}, \emph{i.e.} the step width introduced above. The global minimum was always used as initial starting geometry and local optimizations were performed until the maximum residual force was smaller than 0.025~eV/\AA{}. Local optimizations were performed using a standard quasi-Newton optimizer and Cartesian coordinates.

\section{Results and Discussion}
\label{results}

In our recent work we have already described the efficiency of the DIC-based structure search approach for metallic and hydrogenated silicon clusters in gas phase and when adsorbed on surfaces.~\cite{Panosetti2015} In the following we shortly review these results in the present context and proceed to evaluate the efficiency and applicability of structure search based on DIC (CDIC) trial moves for global optimization of molecules in gas phase and adsorbed on surfaces.

\subsection{Performance of Delocalized Internal Coordinates for Isolated Molecules and Clusters}
\label{results-gasphase}

In our previous work~\cite{Panosetti2015} we have compared the efficiency of BH when using CCs and DICs as trial moves by performing a series of 500-step runs with different parameters (step width, percentage of DICs) for a hydrogenated silicon cluster Si$_{16}$H$_{16}$ both in isolation and adsorbed on a Si(001) surface. Regardless of the choice of parameters, all the runs in DICs were shown to sample a more relevant region of the PES, that is, the minima identified with DIC trial moves are more often intact and the distribution of their energies is invariably centered around lower energies compared to the corresponding CC run with equivalent parameters. Furthermore, CC-based runs almost always failed to identify the global minimum, whereas almost all DIC-based runs found it. Finally, as DIC trial moves produce less strained geometries, the local relaxation at each global step necessitates on average 30\% fewer local optimization steps, which drastically reduces the overall simulation time.

We ascribe the superior performance of DICs to their collective account of the structure and chemistry of the system. DICs by construction take into account the connectivity of atoms to produce collective displacements that resemble vibrational modes. Such displacements are largely composed of concerted motions of groups of atoms, thus naturally suitable to explore configurational subspaces in which certain bonding patterns are preserved. The latter condition is particularly desirable in the investigation of both the gas-phase and the adsorption behavior of large functional molecules, where two intertwined aspects need to be addressed: {\em i)} the structural integrity is a necessary prerequisite to enable functionality; {\em ii)} even small conformational changes, or minor isomerizations, can determine dramatic variations of electronic properties ({\em e.g.} of molecular switches). The structural screening of molecules of this kind should therefore be constrained to subspaces of the PES that satisfy these aspects, while ideally retaining a sufficient degree of flexibility to be able to capture non-intuitive, potentially relevant geometries. 

 \begin{figure}
 \includegraphics[width=\columnwidth]{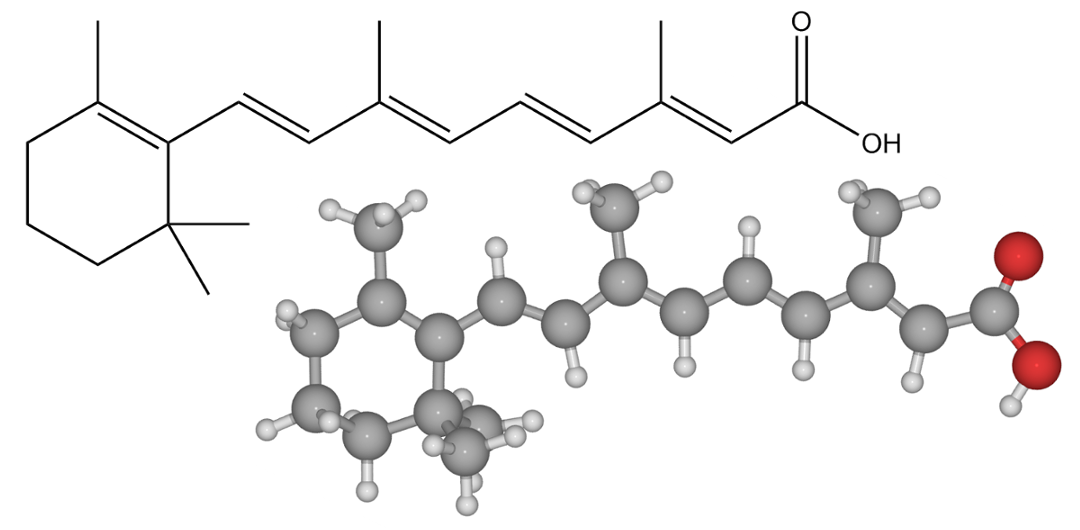}
\caption{\label{fig-gas-pics} Chemical formula (top) and ball-and-stick representation (bottom) of isolated all-trans retinoic acid (ReA). Carbon atoms are colored gray, hydrogen atoms are colored white, and oxygen atoms are colored red. The global minimum geometry in gas phase is shown.}
\end{figure}

In the case of large organic molecules, we need to additionally account for the fact that their degrees of freedom show largely varying bond stiffness. High-frequency stretch motions will for example generally not be relevant for a conformational structure search. To test the efficiency of the DIC-BH approach in such cases, global optimization of isolated ReA (\emph{cf.} Fig.~\ref{fig-gas-pics}) has been performed using 100 trial moves based on random CCs and displacements along varying subsets of DICs. Table~\ref{tab-CC-vs-DIC} compiles the number of symmetry-inequivalent minima found. It is evident that random CCs are particularly ill-suited for large organic molecules. In most of the cases the molecule ended up dissociated and once the molecule is torn apart, regaining an intact structure by random CC displacements is highly improbable. For most applications involving organic molecules, dissociated structures are not of interest and the computing time used for the local optimization step is essentially wasted. When using random CCs, this can amount to up to 95\,\% of the overall computing cost.

\begin{table}
\caption{\label{tab-CC-vs-DIC} Number of symmetry-inequivalent, intact minima found when performing 100 global optimization steps for isolated ReA with step width \emph{dr}, and displacements in CCs or different amounts of DICs and DICs, where the bond stretches have been constrained (constr.\,DICs). Since not a single intact structure was already found for CC displacements with a step width of 0.70, larger step widths have not been considered.}
 \begin{tabular}{lC{1cm}C{1cm}C{1cm}C{1cm}} \hline
            &  \multicolumn{4}{c}{step width $dr$} \\
  \multicolumn{1}{l}{coords.}   & 0.55 & 0.70 & 0.90 & 1.20  \\ \hline
  CC        & 1  & 0  & -  & -  \\
  DIC 10\,\%  & 2  & 2  & 3  & 3  \\
  DIC 25\,\%  & 5  & 2  & 2  & 1  \\
  DIC 100\,\% & 1  & 3  & 2  & 1  \\\hline
 \multicolumn{1}{l}{coords.}    & 0.55  & 0.70  & 1.50  & 2.00  \\ \hline 
  constr.\,DIC 10\,\%  & 2 & 3  & 8  & 8  \\ \hline
 \multicolumn{1}{l}{coords.}    & 0.70  & 0.90  & 1.50  & 2.50   \\ \hline 
  constr.\,DIC 25\,\% &2 &4 &5 &9  \\ 
  constr.\,DIC 100\,\% &2 &6 &4 &9 \\ \hline
 \end{tabular} 
\end{table}

The use of DIC displacements already greatly improves on this. Steps that lead to dissociations can be reduced to 75\,\% of the overall amount of steps, while still finding more relevant minima compared to random CCs. The code responsible for creating the ICs automatically detects dissociations, such that they can be discarded directly at runtime prior to local optimization. This shows that DICs can already be useful in their most general form. However, for an efficient search of the conformational phase space of a large organic molecule this can even further be improved by removing bond stretches as explained in Chapter~\ref{methods}. When the bond stretches are fully constrained (constr. DICs in Table~\ref{tab-CC-vs-DIC}), a much finer screening of the PES is possible, and more unique and intact minima are found. Many different conformers (for example eclipsed/staggered stereochemistry) can be observed which would be extremely difficult to find using trial moves based on CCs or unconstrained DICs. Dissociation events can be reduced to as low as 33\,\% of all steps even at large step widths.

While in the simple example of ReA global optimization might not be necessary to identify the most relevant conformations, these results demonstrate that customizing DICs with structure-preserving constraints can make them more applicable to covalently-bound systems. We conclude with a note on the step width and percentage of DICs. Due to the normalization, the result of the algorithm is highly dependent on the percentage of DICs used. The more DICs a displacement is made up of, the less influence a single DIC has on the overall displacement. Individual DICs contribute more to the 10\% than to the 25\% DIC displacements. To find certain minima some atoms might have to be moved over a few \AA{}ngstr\"om, so a large step width would be of advantage. However, for our test system CC displacements failed to find a single minimum already for a step width as small as 0.70. For DIC displacements the step width can and should be chosen much higher than that. Optimization runs using constr.\,DIC displacements performed best for a step width of above 2.00.

\subsection{Performance of Delocalized Internal Coordinates for Molecules on Surfaces}
\label{section-beta}

We proceed to apply the here presented approach to two molecules adsorbed on surfaces, a $\beta$-acid molecule on Au(111) and methane on Ag(111). The first case exhibits a large number of internal degrees of freedom, the second a shallow PES regarding translations and rotations on the surface.

\subsubsection{\emph{trans}-$\beta$-ionylideneacetic acid adsorbed on Au(111)}
\label{section-ReA-Au}
\begin{figure}
 \includegraphics[width=1.\columnwidth]{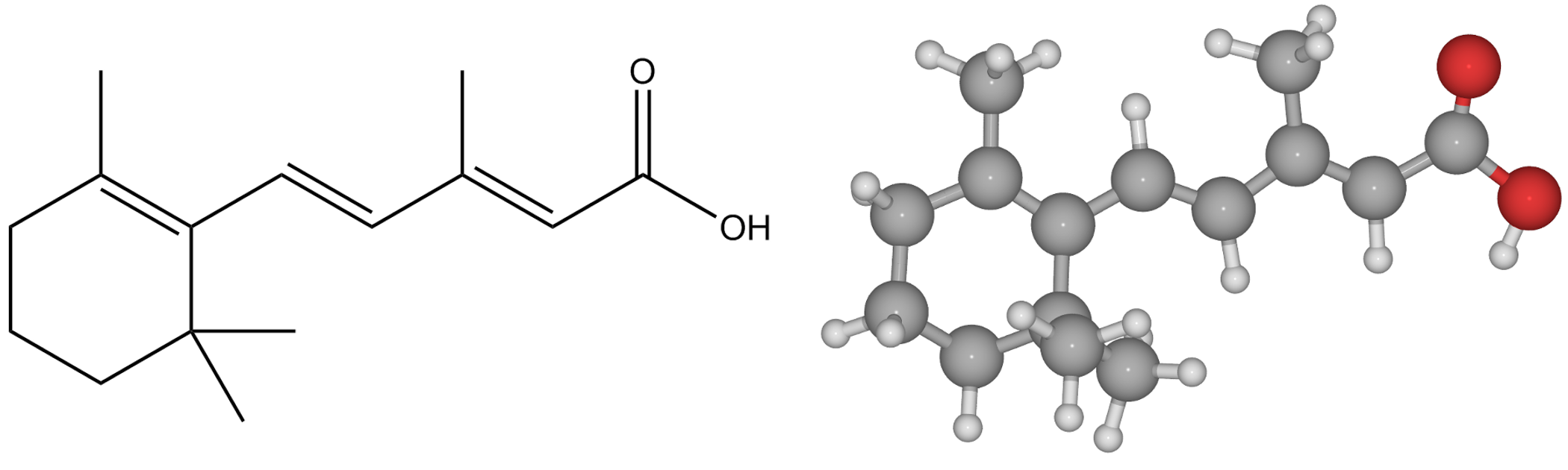}
\caption{\label{fig-beta}Chemical formula (left) and ball-and-stick representation (right) of \emph{trans}-$\beta$-ionylideneacetic acid ($\beta$-acid). Carbon atoms are colored gray, hydrogen atoms are colored white, and oxygen atoms are colored red. The global minimum geometry in gas phase is shown.}
\end{figure}

Even though there might be crystallographic data for large biological molecules, the current resolution of experiments is often insufficient to resolve the individual atomic positions of larger functional molecules adsorbed on surfaces. Closely related to their role in nature, functional molecules could act as molecular switches when adsorbed on a metal surface.~\cite{Barth_review_2007,Karan2015, Maurer2012} Therefore, prediction of key structural elements by electronic structure methods is crucial. We study the performance of CDIC-based global optimization on the example of a smaller analogue of ReA, namely \emph{trans}-$\beta$-ionylideneacetic acid~\cite{Shirley1985} ($\beta$-acid, \emph{cf.} Fig.~\ref{fig-beta}) adsorbed on a Au(111) surface. As an initial starting geometry, the molecule in the conformation representing the global minimum in gas phase was placed flat lying at a random adsorption site on the surface, and the structure was relaxed using local optimization. No surface relaxation was allowed. This was used as a starting point for all BH runs. As a reference, 100 global steps were performed for step widths of 0.25, 0.5, and 1.0 using CC trial moves. This is compared to 100 global step runs using 10\%, 25\% and 100\% mixtures of constr.\,CDICs, all with a step width of 1.5.

For surface-adsorbed molecules, the shortcomings of standard CC displacements become even more apparent as shown in Table~\ref{tab-CC-sur}. In 70--90\,\% of all global optimization steps, a dissociated structure is obtained. Most commonly the cyclohexenyl ring is torn apart or the conjugated side chain is broken. In the context of conformational switching, this would render this minimum irrelevant for the process under study. Keeping important structural motifs intact is difficult with CC displacements. One could argue that setting a smaller step width remedies this problem; however, our results show that, as the step width decreases, the starting geometry is found with higher probability. There seems to be no optimal setting for the step width parameter: either the molecule dissociates, or the sampling is not able to escape the initial basin of attraction. As a result, we did not obtain any single new geometry other than the starting point and there was almost no lateral adsorption site sampling. CC trial moves are thus not applicable to this case. 

\begin{table}
\caption{\label{tab-CC-sur} Number of dissociated structures, revisits of the starting geometry, sampling of new adsorption sites in the conformation of the gas-phase minimum, and new structures when performing 100 global optimization steps for $\beta$-acid on Au(111) with different step widths using CC trial moves.}
 \begin{tabular}{lC{1cm}C{1cm}C{1cm}} \hline
     	&  \multicolumn{3}{c}{step width $dr$} \\
         \multicolumn{1}{c}{CC trial moves}    	&0.25&0.5&1\\\hline
Dissociations&73&93 &98 \\
Starting geometry&24 &7 &2 \\
Different adsorption site&{3} &{0}&{0} \\
New structures&{0} &{0} &{0}  \\\hline
\end{tabular}
\end{table}

\begin{table}
\caption{\label{tab-DI-sur} Same as Table~\ref{tab-CC-sur}, but when using displacements in constr. CDICs (denoted by the percentage of constr.\,CDICs randomly selected) with a step width $dr$ of 1.5.}
 \begin{tabular}{lC{1cm}C{1cm}C{1cm}} \hline
&  \multicolumn{3}{c}{amount of constr.\,CDICs} \\
   \multicolumn{1}{c}{constr.\,CDIC}     		&10\,\%&25\,\%&100\,\%\\\hline \noalign{\vskip 2pt} 
Dissociations &62 &38 & 37\\
Starting geometry &8 &3 &30 \\
Different adsorption site&19 &{48} &{23} \\
New structures&{11} &11 &{10} \\\hline
\end{tabular}
\end{table}

\begin{figure*}
\centering
 \includegraphics[width=\textwidth]{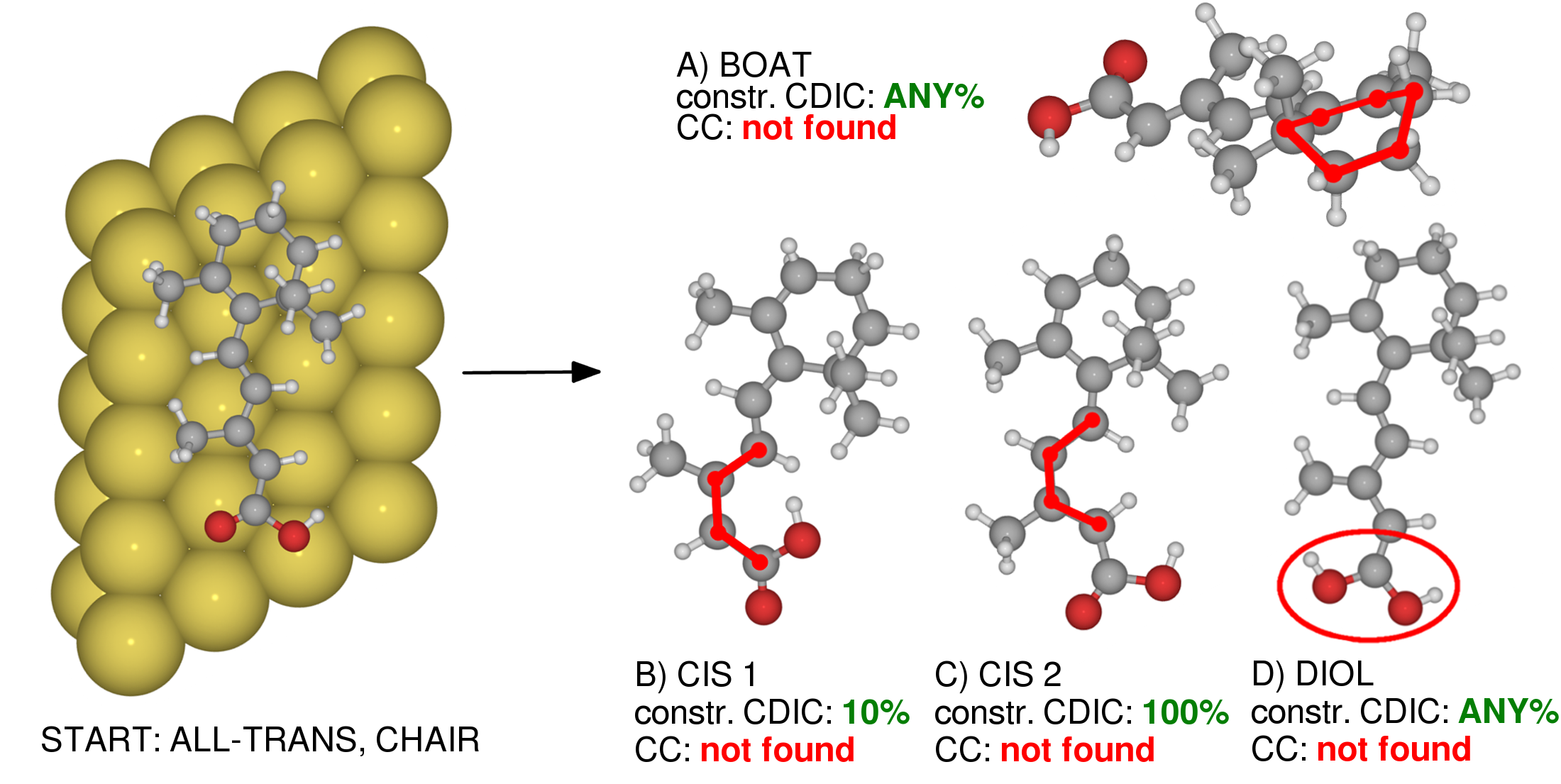}
\caption{\label{fig-strucis}Ball-and-stick representations of the starting geometry (left), of the boat configuration (top right, A), of the \emph{cis} isomers (B and C), and of the diol structure (bottom right, D) of $\beta$-acid at Au(111). All were found using constr. CDIC displacements in less than 100 global optimization steps. The \emph{cis} configurations as well as the ring in the boat configuration have been marked in red for clarity. Also shown are the settings in CDIC runs, at which the geometries have been found. None of these structures were found using CC trial moves.}
\end{figure*}

The result of the sampling is changed dramatically by the introduction of CDIC trial moves with constrained stretches (constr.\,CDIC, \emph{cf.} Table~\ref{tab-DI-sur}). Using 25\% and 100\% of available constr.\,CDICs reduces dissociation events to about 40\% of all steps. For 10\% constr.\,CDIC displacements at a step width of 1.5 the amount of dissociations is comparable to CC trial moves using a step width of 0.25, however, at largely different sampling efficiency. When fewer DICs are mixed into a displacement, the influence of each individual DIC in the final trial move is larger due to normalization. As a consequence and consistent with the gas-phase sampling results above, 10\,\% constr.\,CDIC displacements introduce the most significant changes to the molecular geometry, {\em i.e.} result in more dissociations. Inversely, 100\% constr.\,CDIC displacements sample the starting geometry more often than 10\% or 25\%  constr.\,CDIC trial moves. As the number of translations and rotations in comparison to the number of internal degrees of freedom in the overall displacement is smaller at higher percentages of CDICs, lateral sampling is less efficient than when using 25\% or 10\% constr.\,CDIC trial moves. It is not straightforward to find the optimal position and orientation of a complex adsorbate like the one studied here, and efficient lateral sampling is therefore strongly desired. In contrast to standard CC based trial moves, different rotations and various adsorption sites of the molecule have been sampled during all three runs. A closer analysis of this aspect will be presented in section~\ref{section-methane}. 

Conformational structure search for $\beta$-acid on Au(111) serves the purpose of finding new and stable configurations. Especially interesting are {\em cis} isomers (\emph{cf.} Fig.~\ref{fig-strucis}, structures B and C) since they could play an active role in a molecular switching process. These structures are particularly hard to uncover by unbiased sampling, since a concerted motion of a large number of atoms is necessary. We were able to find such minima in two out of three constr.\,CDIC based runs (10\% and 100\% constr.\,CDIC trial moves) with a relatively short number of steps per run (25 and 88 global steps, respectively). It should be noted that CC based sampling, however, was not able to uncover any new conformations apart from the starting geometry.

Other minima of interest are conformations of the molecule that might all contribute to the apparent finite-temperature geometry observed in experiment, such as different ring conformations (chair, boat, half-chair etc.) or changes in internal degrees of freedom (chain to ring angle, staggered~vs.~eclipsed conformations of methyl groups etc.). The broadest range of such structures has been identified using 10\% constr.\,CDIC displacements and a step width of 1.5. Configurations were found where the ring was bent towards the surface and away from it. We have also found stuctures where the ring is in a boat configuration (\emph{cf.} Fig.~\ref{fig-strucis}, structure A) or twisted with respect to the chain. Methyl groups both on the ring and on the chain were found in different rotations (staggered conformation). Another group of minima involved modifications of the acid group, rotating either the OH or the whole COOH group by 180$^\circ$. 25\% constr.\,CDIC trial moves found similar minima, but were not able to sample changes in the acid group. Instead, the boat configuration of the ring was sampled more often. The third setup (100\% constr.\,CDIC displacements) did not result in a completely rotated acid group, the OH group was found flipped once. This run was the only one that sampled half-chair conformations of the ring. A somewhat unexpected result was the repeated discovery of a diol structure (\emph{cf.} Fig.~\ref{fig-strucis}, structure D) in all constr.\,CDIC runs, while CC based runs always failed to find this structure. We have performed DFT calculations that predict a stabilizing effect of the surface by 0.22~eV on this structure as compared to the gas-phase minimum conformation; however, the acid structure nevertheless remains the global minimum (see Appendix~\ref{appendix-dft}). The energetically most favorable structures that were found by all runs are slight variations of the global minimum conformation found in gas phase, adsorbed such that the oxygen atoms are situated above a bridge site.

In summary, CC displacements were found to either repeatedly sample the starting geometry at small step widths or dissociated structures at larger step widths. Using CC trial moves we were not able to find any step width that is both large enough to overcome barriers on the PES and simultaneously small enough not to dissociate the molecule. Expressing the displacements in constr. CDICs, however, enables a finely tunable sampling of phase space. Indeed a small number of steps are sufficient to find a plethora of molecular configurations on the surface as well as different adsorption sites and rotations.

\subsubsection{CH$_{4}$ adsorbed on Ag(111)}
\label{section-methane}

\begin{figure}
\centering
 \includegraphics[width=\columnwidth]{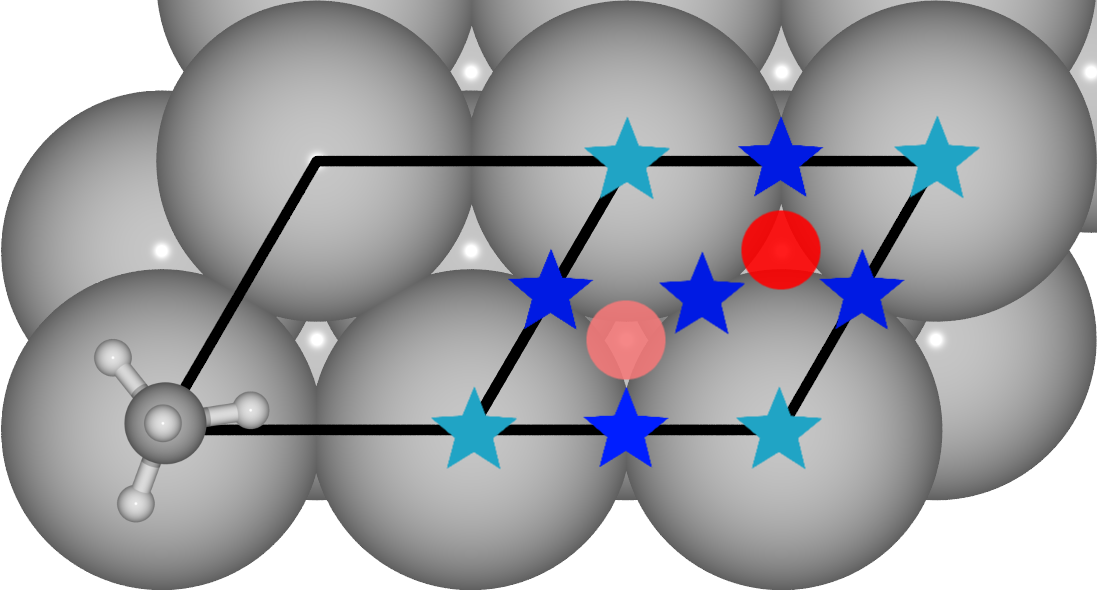}
\caption{\label{fig-fcc} Methane (CH$_4$) adsorbed on Ag(111). Shown are two neighboring primitive surface unit cells with the initial starting geometry used in the sampling runs on the left hand side, while high-symmetry adsorption sites are highlighted on the right hand side: fcc hollow site (light red circle), hcp hollow site (dark red circle), top sites (light blue star), bridge sites (dark blue star). All top sites, as well as all bridge sites, are symmetry equivalent.}
\end{figure}

The example of $\beta$-acid on Au(111) illustrated the importance of chemically-adapted coordinates in sampling the internal degrees of freedom of an adsorbate on a surface. To better understand the sampling of combined surface/adsorbate degrees of freedom such as the lateral adsorption site and the orientation of the molecule on the surface we next study CH$_4$ (methane) adsorbed on Ag(111). Different adsorption sites of CH$_4$/Ag(111) are found to have similar stability and the PES is dominated by small barriers and little corrugation between these sites. As a consequence, it should be an ideal test system for which we can sample all adsorption geometries in the unit cell. 

Even though the system appears simpler than the previous test cases, a large number of distinct minima exists. The surface has four different high-symmetry adsorption sites (\emph{cf.} Fig.~\ref{fig-fcc}) and although the molecule is highly symmetric, CH$_4$ can adsorb in multiple rotational orientations with respect to the surface. In principle, most of these minima might be found by visual inspection; however, as emphasized by Peterson,~\cite{Peterson2014} as soon as more than one adsorbate is introduced or if the surface features steps or defects, combinatorics makes brute-force sampling and analysis methods quickly intractable. We performed global optimization runs with Cartesian displacements (step width 0.4) as well as constr. CDIC displacements made up of 25\% and 75\% of the available coordinates (step width 1.25) until 260 intact CH$_4$ minima were found. This took 600 global optimization steps with CC trial moves and 500 with constr.\,CDIC trial moves, respectively. Step widths were chosen according to the best performance in shorter test runs (100 steps each). As initial starting geometry, the molecule has been positioned at a top site with one hydrogen atom pointing away from the surface, followed by local optimization (\emph{cf.} left hand side of Fig.~\ref{fig-fcc}). Optimization results have been obtained for CH$_4$ in a (2x2) unit cell to mimick a lower coverage at the surface.

\begin{figure}[htb]
\centering
 \includegraphics[width=\columnwidth]{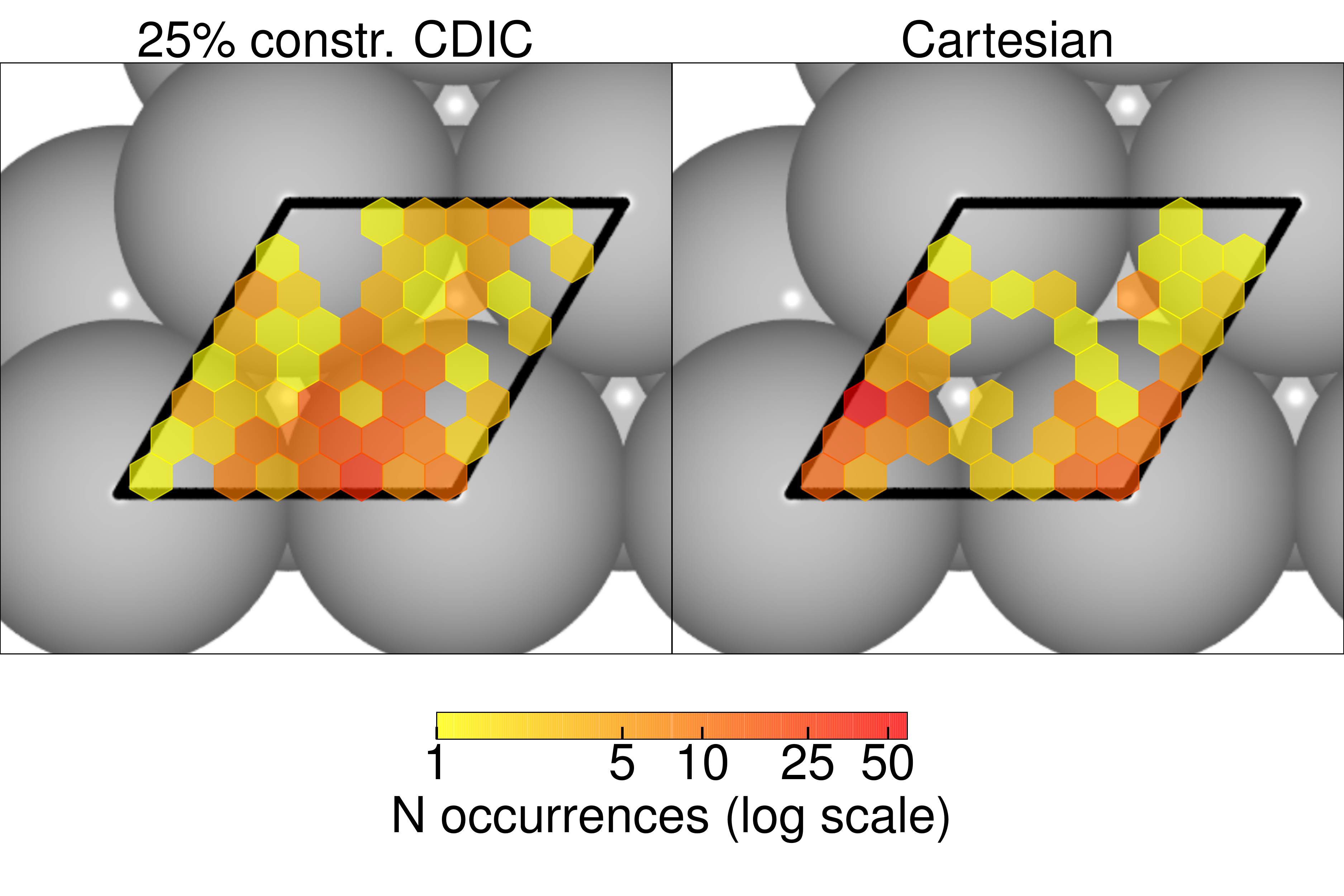}
\caption{\label{fig-2dhist}2D histogram of the $x$,$y$-position of the carbon atom in CH$_4$ in all 260 intact methane minima identified over the course of the 25\% constr.\,CDIC displacements (left side) and CC displacements (right side) sampling run (see text). Lighter yellow hexagons indicate a rare sampling of this position in the surface unit cell, darker orange or red hexagons are shown when this position was sampled more often. No hexagons are drawn when the simulation failed to ever visit this area on the surface. Calculations were performed for a (2x2) unit cell to mimick lower coverage, but folded back into a  primitive unit cell as indicated by black lines.
}
\end{figure}
% Constr.\,CDIC displacements rigorously sample top, bridge and both hollow sites, while CC displacements get preferentially stuck close to the top adsorption sites and fail to properly sample one of the hollow sites.

To achieve a rigorous sampling on a metal surface, the first prerequisite is a complete lateral exploration of the surface unit cell. In our test case, this means that the simulation has to traverse through all four distinct high-symmetry adsorption sites (right hand side of Fig.~\ref{fig-fcc}). We therefore monitor the position of the carbon atom in all intact methane minima identified over the course of the sampling run and plot the result in a 2D histogram in Fig~\ref{fig-2dhist}. It immediately stands out that CC trial moves fail to sample larger areas of the surface unit cell. This includes rare sampling of the bridge sites, the center of the fcc hollow site, as well as the surroundings of the hcp hollow site. Top sites are instead sampled repeatedly, and computational sampling time is wasted in these revisits. In the case of constr.\,CDIC trial moves the distribution of identified minima is more homogeneous and all four high-symmetry sites are visited. An especially interesting finding is that the most often sampled areas are not around the starting site, but around the bridge sites. There is also a slight preference for the fcc hollow site. Since translations are explicitly included in constr.\,CDIC displacements on surfaces, it is not a big surprise that our trial moves perform better in this regard. In contrast, a random CC trial move is unlikely to describe a concerted molecular translation across the surface.

The second characteristic of a thorough structural sampling in this case is the variety of rotational arrangements of the methane molecule that are found. CC trial moves mostly produce geometries close to the top sites with one hydrogen atom pointing towards the center of the corresponding surface silver atom. A minimum with two hydrogen atoms wedged in-between the bridge site, the others pointing towards the top sites is found twice. The most stable configuration, which is only slightly more favorable than other minima, corresponds to adsorption at the hcp hollow site, with the hydrogen atoms rotated to lie above the ridges between silver atoms (staggered). This geometry at the hcp and fcc hollow site was found in the BH run based on CC trial moves. The constr.\,CDIC trial moves were instead able to find all these rotational arrangements, as well as a number of others. Both runs (25\% and 75\% constr. CDIC) sampled a variety of different orientations at the bridge sites and close to the top sites as well. The 75\% constr.\,CDIC run even found the molecule flipped around completely once, with one of the hydrogen atoms pointing towards the surface. 25\% constr.\,CDIC trial moves were the only ones that produced an eclipsed rotational configuration of hydrogen atoms at both hollow sites, with the H atoms oriented along the close-packed rows of the Ag(111) surface.

\begin{figure}[htb]
\centering
 \includegraphics[width=\columnwidth]{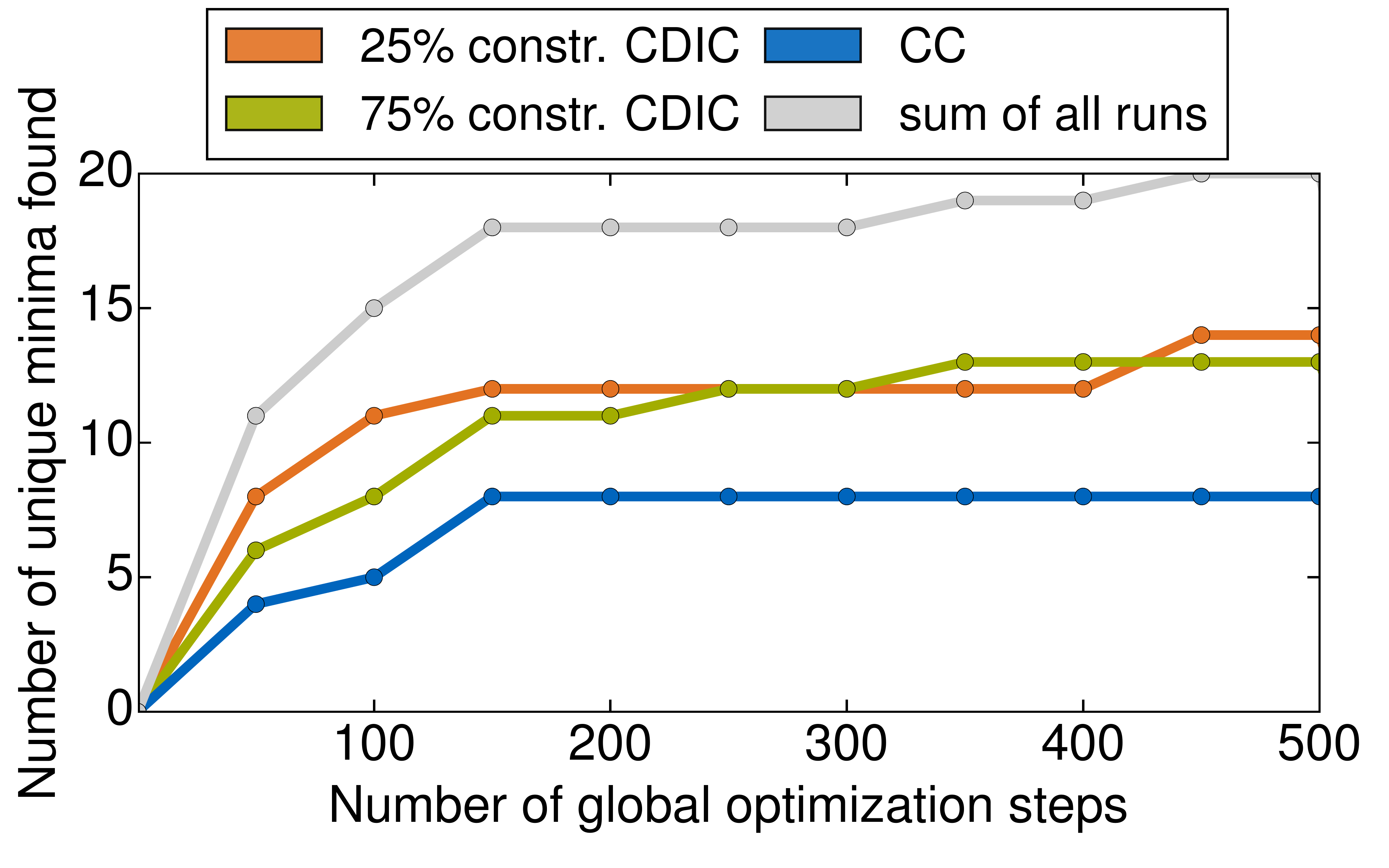}
\caption{\label{fig-unique}20 symmetry-inequivalent positions and orientations of CH$_4$ on Ag(111) were found in the two CDIC (25\% and 75\%) and one CC displacement based BH global optimization runs combined. Depicted is the number of these minima found in each run up to a certain number of global steps. All minima that were encountered during the CC displacement based search were also discovered in at least one of the two constr.\,CDIC runs.}
% \todo[color=green,inline]{KK: Ich hatte einen Fehler ins Skript eingebaut als ich das von absoluter Minimazahl zu Prozent umgestellt hab. 8 Minima werden kartesisch gefunden was etwa die Hälfte von den CDIC runs ist. Ich nehme an bei irgendeiner revision ist daraus die Hälfte aller Minima, also 10, geworden.}
\end{figure}

Fig.~\ref{fig-unique} shows how the discovery of new, unique CH$_4$ minima occurred over the course of the global optimization runs. In all three runs combined, a total of 20 symmetry-inequivalent minima was found, of which the CC based run was only able to find 8. All these minima were also identified in either one or both of the constr.\,CDIC runs. An interesting finding here is that the constr.\,CDIC runs found twice as many inequivalent geometries as the CC run already after about 100 global optimization steps. After an initial phase of 150 steps, CC trial moves do not lead to any more new minima, whereas constr.\,CDIC runs continue to find more. Both effects together clearly show the superior performance of constr.\,CDIC trial moves compared to simple CC trial moves.

\begin{figure}
\centering
 \includegraphics[width=\columnwidth]{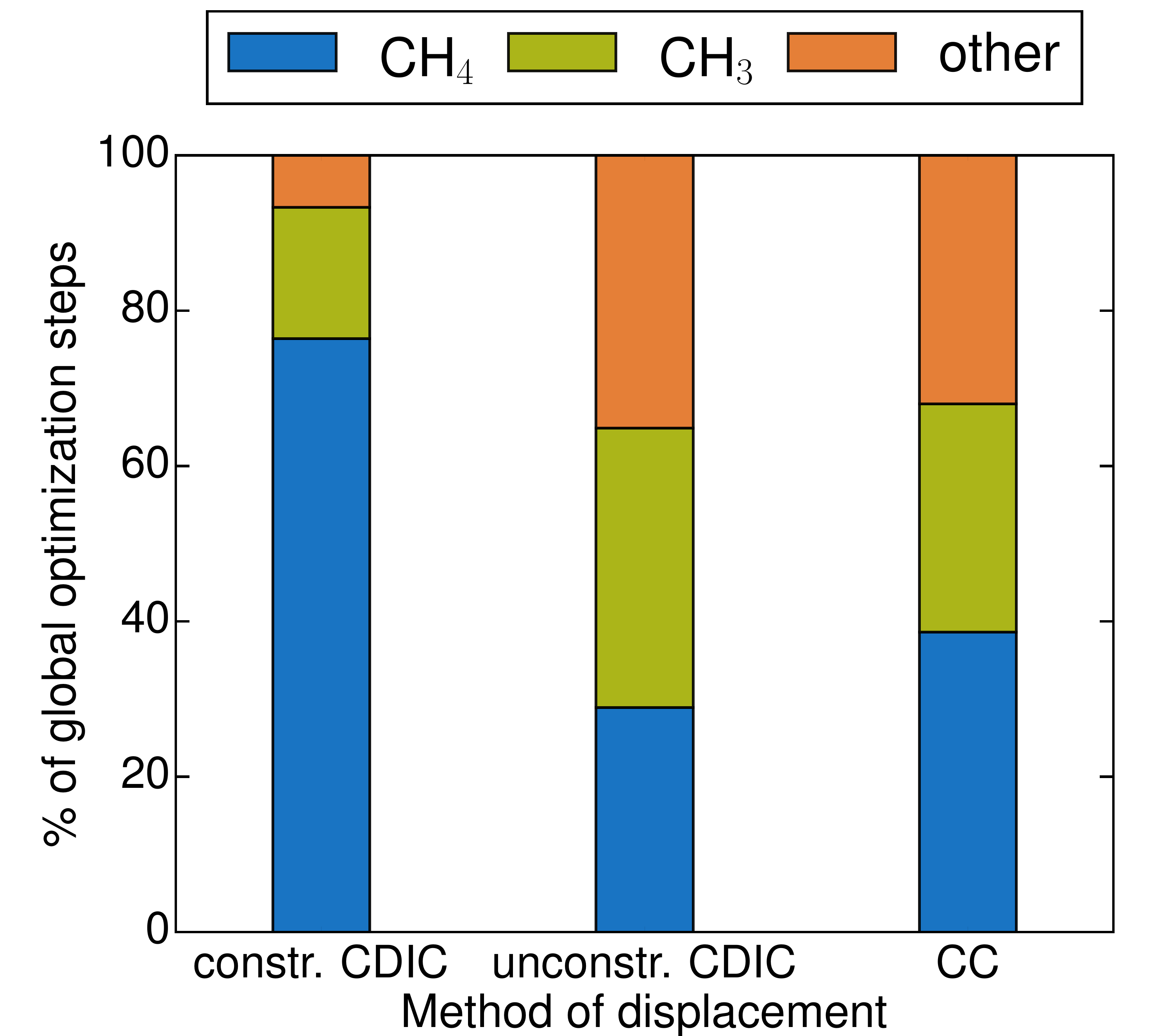}
\caption{\label{fig-stacked}Distribution of all minima found using different displacement methods for the first 500 global optimization steps of each run. Shown are results for 25\% constrained (step width 1.25) and unconstrained CDIC runs (step width 1.6) and the CC run (step width 0.4).
Intact CH$_4$ structures are shown in blue, CH$_3$ in green, and all other minima (CH$_2$, CH and C) as orange.}
\end{figure}

Up to this point we have focused on geometries where the CH$_4$ molecule remained intact. However, CDIC trial moves are equally able to efficiently sample the space of chemically different minima that involve hydrogen dissociations and chemical reactions on the surface. CC displacements are completely defined by the single parameter of step width $dr$, and there is no straightforward way of fine tuning the search target, for example in terms of focusing on mainly dissociated or mainly intact geometries of CH$_4$. CDIC displacements, however, enable a straightforward inclusion of constraints that confine configurational space, for example by constraining bond stretches. In order to analyze the distribution of dissociated and intact structures throughout a BH run, we additionally performed 500 steps of an unconstrained CDIC run using 25\% of available coordinates and a step width of 1.6. Figure~\ref{fig-stacked} shows the distribution of intact and dissociated minima for different displacement methods. As also found in the case of $\beta$-acid on Au(111), constr.\,CDIC displacements effectively restrict the geometry search to intact CH$_4$ molecules. More importantly, unconstrained CDIC trial moves find a similar distribution of dissociated and intact structures as CC displacements and are therefore equally suited to sample the space of reactive intermediates at the surface. However, this comes at a higher sampling rate of unique minima than achieved with CC trial moves.

%%KR Die nächsten Zeilen und die Abbildung verstehe ich nicht. Bisher habt ihr nur 3 runs erklärt: 25% und 75% CDIC und einen CC run. Was wird hier jetzt diskutiert, wenn von constrained CDIC und "`nur"' CDIC gesprochen wird? Was soll in der caption "`at least"' 500 Steps bedeuten? Das muss alles klar eingeführt und definiert werden! Was bedeutet im letzten Satz unten "`at a higher sampling rate"'?
% \todo[inline,color=green]{KK: comp det. Methane (CH$_4$) on 4 layers of a ($2 \times 2$) Ag (111) surface unit cell was investigated with a step width of 0.40 using CC displacements and 1.25 when using 25 and 75\,\% of available constr.\,CDICs. Additionally a run using 25\,\% unconstrained CDICs was performed with a step width of 1.6. Each set of settings was tested for 500 global optimization steps. The maximum residual force for local optimizations was 0.025~eV/\AA{}. }

Lastly, there is one more mechanism with which DICs increase the computational efficiency of global optimization. The most expensive part of a global optimization step is the local optimization, which typically uses a large number of steps to converge to the local minimum energy structure. Reducing the average number of such local optimization steps can lead to a significant computational speed-up. For all test cases we find that constr.\,CDIC trial moves generally generate geometries that are less strained compared to the ones produced by CC trial moves. This is emphasized by the fact that geometries created from CC trial moves take almost twice as many steps for local convergence as ones generated by 75\% constr.\,CDIC displacements, namely on average 528  BFGS~\cite{Broyden1970,Fletcher1970,Goldfarb1970,Shanno1970} steps compared to 267 using a force threshold of 0.025~eV/\AA{}.
As explained in section~\ref{section-ReA-Au}, using smaller percentages of constr.\,CDICs introduces more severe changes to the molecule. Therefore 25\% constr.\,CDIC displacements take slightly longer for convergence as 75\% constr.\,CDIC trial moves (309 steps). Overall this nevertheless constitutes a significant reduction in computational cost, consistent with what we already observed in our previous work on silicon clusters.~\cite{Panosetti2015} In summary, we thus find that global structure search with DIC trial moves not only finds more unique minima than CC displacements, but does so at reduced computational cost as well.

\section{Conclusions and Outlook}
\label{conclusions}

We presented a modification of the basin hopping algorithm by performing global optimization trial moves in chemically-motivated curvilinear coordinates that resemble molecular vibrations. This approach has recently been shown to be efficient in structure determination of clusters~\cite{Panosetti2015} and was here investigated and extended for the application to organic molecules both in gas phase and adsorbed on metal surfaces. The chemical nature of the collective displacements enables straightforward inclusion of rotations, translations, as well as constraints on these and arbitrary internal degrees of freedom. This allows for a finely tunable sampling of practically relevant areas of phase space for complex systems that appear in nanotechnology and heterogeneous catalysis, such as hybrid organic-inorganic interfaces,~\cite{Maurer2016Review} organic crystals, self-assembled nanostructures, and reaction networks on surfaces. These systems feature heterogeneous chemical bonding, where stiff energetically favorable bonding moieties, such as covalent molecules, coexist with weak and flexible bonding forces such as interactions between molecules or between a molecule and a surface. 

For the show cases $\beta$-acid adsorbed on Au(111) and methane adsorbed on Ag(111) we find that collective internal coordinate based trial moves outperform Cartesian trial moves in several ways: \emph{i)} CDICs systematically identify a larger number of structures at equal number of global optimization steps, thus allowing to reduce the number of necessary global optimization steps; \emph{ii)} constr.\,CDICs are able to restrict structure searches to well defined chemically-motivated subdomains of the configurational space and thus find more relevant structures; \emph{iii)} DICs generate less strained structures and reduce the computational cost associated with a single global optimization step.
 
In future work we plan to utilize the variability of singular-value decomposed collective coordinates in the context of materials structure search to facilitate simulation of surface reconstructions and defect formation, as well as crystal polymorphism and phase stability in organic crystals and layered materials. So far we have only discussed the relevance of these coordinate systems in the context of basin hopping, however displacement moves are a common element of many different stochastic global optimization strategies, such as, for example, minima hopping, where curvilinear constraints can help guide the MD.~\cite{Peterson2014} Our implementation can easily be extended to feature such algorithms as well. 

\appendix

\section{\texttt{winak}: a Python-based tool to construct and apply curvilinear coordinates for materials structure search and beyond}
\label{appendix-winak}
Many global optimization procedures can be divided into three steps: displacement, optimization, evaluation and analysis. In order to make DIC displacements as easily accessible as possible, we have developed a modular Python-based framework called \texttt{winak},~\cite{winak} where each of the three aforementioned steps can be customized. For instance, in order to change from the basin hopping method to the minima hopping method~\cite{Goedecker_JCP_2009}, instead of a global optimization, an MD run can be started, and instead of employing a Metropolis criterion, all newly identified minima are accepted.~\cite{Peterson2014} 

In order to make the code independent of a particular application, each part of the global optimization process has been encapsulated in an individual class. Exchanging local optimization by an MD run can easily and safely be done, independent of the coordinate generation class. The same is true for the displacement step. Logging and error handling is managed by analysis routines that work with any combination of sub classes. To ensure compatibility, abstract base classes and templates are provided on the basis of which custom procedures for the three above mentioned steps can be developed. This code is designed to be interfaced with ASE,~\cite{Bahn2002} which enables a wide range of electronic structure codes to be used in combination with \texttt{winak}.

\section{Density-Functional Theory calculations}
\label{appendix-dft}

We repeatedly encountered diol isomers in the case of \emph{trans}-$\beta$-ionylideneacetic acid adsorbed on Au(111) on the DFTB level. To further investigate this, we recalculated the energy differences between acid and diol, both in gas phase and on the surface using DFT. All calculations were carried out using the Perdew-Burke-Ernzerhof (PBE) exchange-correlation functional,~\cite{Perdew1996} implemented in the FHI-aims code using the light basis set.~\cite{Blum2009,Ren2013} Local optimization was considered converged when the maximum residual force was smaller than 0.025~eV/\AA{}. Calculations including the surface were done on 4 layers of (4$\times$6) Au (111) with a k-grid of $(5 \times 5 \times 1)$. Surface atoms were not allowed to relax during optimization. Both in gas phase and on the surface the acid isomer is more stable, which aligns with chemical intuition. However, the energy difference is lowered from 1.09\,eV in the gas phase to 0.87\,eV for the adsorbed system, {\em i.e.} the diol is stabilized.

\begin{acknowledgments}
The authors would like to thank Daniel Strobusch and Christoph Scheurer for supplying the initial coordinate construction code and for fruitful discussions. RJM acknowledges funding from the DoE-Basic Energy Sciences grant no. DE-FG02-05ER15677. CP gratefully acknowledges funding from the Alexander von Humboldt Foundation and within the DFG Research Unit FOR1282. DP gratefully acknowledges funding from the Engineering and Physical Sciences Research Council (project EP/J011185/1).
\end{acknowledgments}

\raggedright


% \bibliography{references}

\end{document}